\documentclass[journal]{IEEEtran}
\ifCLASSINFOpdf

\else

\fi

\usepackage{graphicx}

\usepackage[cmex10]{amsmath}
\usepackage{amssymb}
\usepackage{cite}
\usepackage{graphicx}
\usepackage{array,color}
\usepackage{multirow}
\usepackage{amsmath}
\usepackage{stfloats}
\usepackage{graphicx}
\usepackage{subfigure}
\usepackage{tabularx}
\usepackage{epsfig,epsf,color,balance,cite}
\usepackage{setspace}
\usepackage{bm}
\usepackage{textcomp}
\usepackage{subfigure}
\usepackage{caption}
\usepackage{graphicx}
\usepackage{setspace}

\begin{document}

\title{Achievable Data Rate for URLLC-Enabled UAV Systems with 3-D Channel Model}

\author{Hong Ren, Cunhua Pan, Kezhi Wang, Yansha Deng, Maged Elkashlan,  and Arumugam Nallanathan, \IEEEmembership{Fellow, IEEE}
\thanks{This work is supported by the U.K. Engineering and the Physical Sciences Research Council under Grant EP/R006466/1 and Grant EP/N029666/1. (\emph{Corresponding author: Cunhua Pan.})}
\thanks{H. Ren, C. Pan, M. Elkashlan and A. Nallanathan are with School of Electronic Engineering and Computer Science, Queen Mary University of London, London, E1 4NS, U.K. (Email:{h.ren, c.pan, maged.elkashlan, a.nallanathan}@qmul.ac.uk). K. Wang is with Department of Computer and Information Sciences, Northumbria University, UK. (e-mail: kezhi.wang@northumbria.ac.uk). Y. Deng is with the Department of Informatics, King's College London, London WC2R 2LS, U.K. (e-mail:yansha.deng@kcl.ac.uk).}
}

\maketitle

\begin{abstract}
In this paper, we investigate the average achievable data rate (AADR) of the control information delivery from the ground control station (GCS) to unmanned-aerial-vehicle (UAV) under a 3-D channel, which requires ultra-reliable and low-latency communications (URLLC) to avoid collision. The value of AADR  can  give insights on the packet size design. Achievable data rate under short channel blocklength is adopted to characterize the system performance.   The UAV is assumed to be uniformly distributed within a restricted space. We first adopt the Gaussian-Chebyshev quadrature (GCQ) to approximate the exact AADR. The tight lower bound of AADR is derived in a closed form. Numerical results verify the correctness and tightness of our derived results.
\end{abstract}

\IEEEpeerreviewmaketitle

\vspace{-0.2cm}
\section{Introduction}
\vspace{-0.1cm}

UAV assisted wireless communication has attracted extensive research attention from both academia and industria \cite{yongzengmaga}, due to their benefits of low cost, swift deployment and high mobility. The link quality between UAV and ground users (GUs) can be enhanced due to the high probability of line-of-sight (LoS) communications.

Most of existing work mainly focused on the study of data transmission links (i.e., UAV-to-GU and GU-to-UAV links), such as energy-efficient trajectory design \cite{yongzeng2017,cunhua}, location optimization \cite{Hourani2014}, and data services in emergency networks \cite{nanzhao,cheng2018}.
However, the control information delivery  from the ground control station (GCS) to the UAV introduce new challenges for UAV communications since these links have stringent latency and reliable requirements for supporting safety-critical functions \cite{yongzengmaga}, such as real-time control to avoid collision. In general, very low data rate is enough to exchange the control information between the GCS and the UAV. In contrast to the conventional communications with relatively long transmission delay  and large packet size, small packet size should be delivered to support the extremely low-latency transmission for control information delivery. In small packet transmission, the Shannon' capacity formula  based on the law of large numbers is no longer applicable, and decoding error probability  cannot be arbitrary small.  In \cite{Polyanskiy2010IT}, the  achievable data rate in finite blocklength regime has been derived. Resource allocation for UAV communication with URLLC was considered in \cite{changyang}.

However, the study on the performance analysis  for  the control information delivery in UAV communications is still missing.  Against this background, we analyze the average achievable data rate for this transmission in UAV communications under 3-D channel model. Two challenges will complicate the analysis: the complicated data  rate expression under short packet transmission and the complicated 3-D channel model.  The contributions of  this paper can be summarised as follows: 1) We are the first to study the average achievable data rate (AADR)  of  control information delivery for a GCS-to-UAV communication system under short packet transmission in 3-D channel model. It can provide engineering insights on the packet size design; 2) The GCQ is adopted to approximate the AADR; 3) The tight lower bound of AADR is derived in closed form; 4) Simulation results verify the correctness and tightness of our derived results.

\vspace{-0.2cm}
\section{System Model}\label{system}
\vspace{-0.1cm}

\begin{figure}[h]
\centering
\includegraphics[width=2.6in]{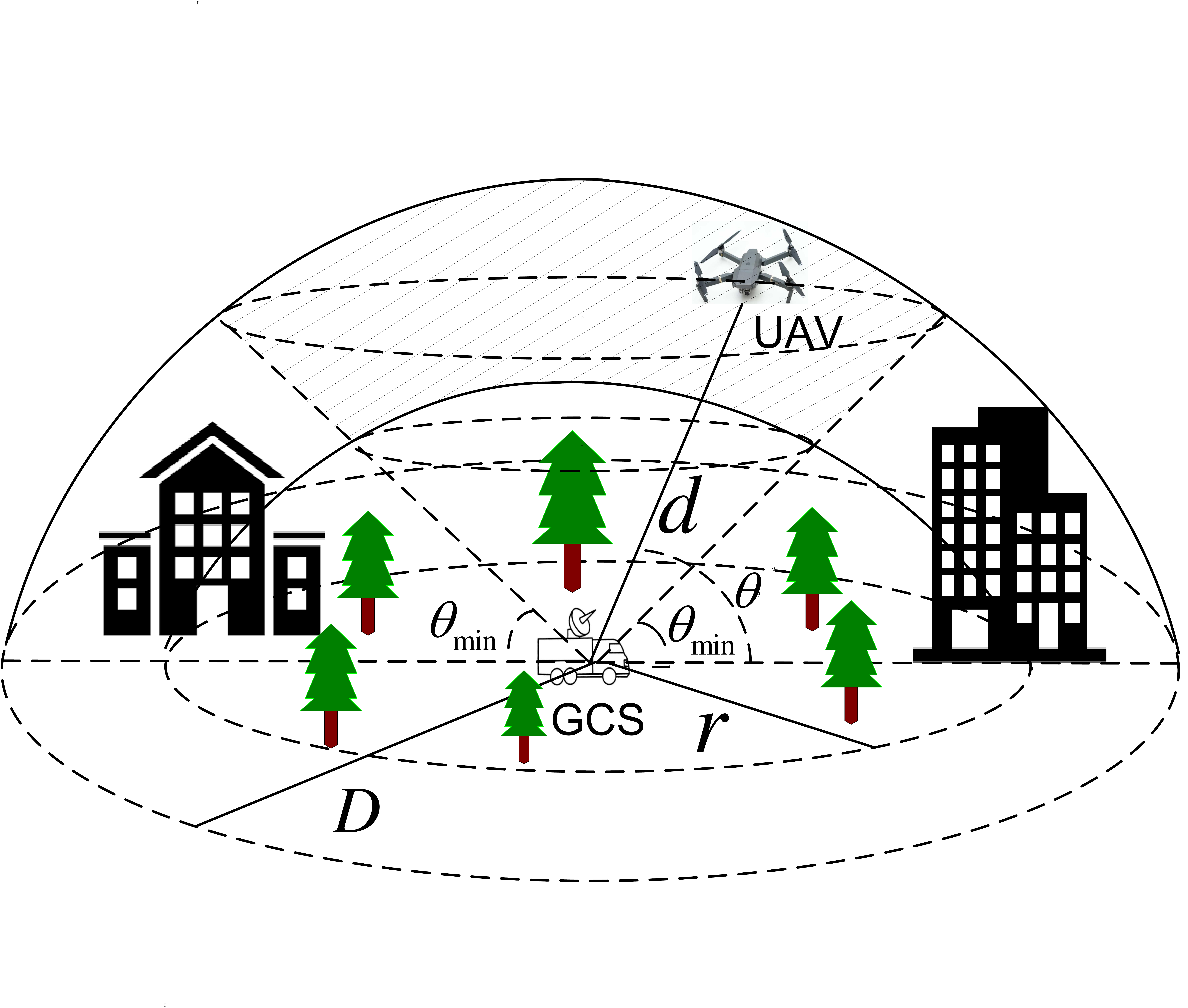}
\vspace{-0.8cm}
\caption{Illustration of the low-latency transmission of control signal from the GCS to the UAV.}
\vspace{-0.2cm}
\label{systemodel}
\end{figure}

As shown in Fig.~\ref{systemodel}, we consider that a GCS sends remote control signals to a UAV, which requires ultra-high reliability and ultra-low latency. Both GCS and UAV are assumed to be equipped with single antenna. Without loss of generality, we assume that the GCS is located at the center of the circle. To guarantee that the UAV is within the GCS's control area, we assume that the UAV is flying within an inverted cone centered at GCC with the largest radius of $D$. We also define a small inverted cone centered at GCS with radius $r$  ($D>r$) that the UAV will not fly into. In addition, to avoid collision onto the buildings or trees around the GCS, the elevation angle of the UAV denoted as $\theta$ should be no smaller than $\theta_{\rm{min}}$ as shown in Fig.~\ref{systemodel}.  The UAV can fly freely in any direction in the space inside the large inverted cone but outside the small inverted cone as shown in the shadow area in Fig.~\ref{systemodel}.
\vspace{-1.1cm}
\\
Then, the cumulative distribution function (CDF) of distance $d$ is
\begin{equation}\label{ww2}
{F_d}(x) = \frac{{{x^3} - {r^3}}}{{{D^3} - {r^3}}},\ r \le x \le D
\end{equation}
and the probability distribution function (PDF) is
\vspace{-0.1cm}
\begin{equation}\label{jijioijojio}
\vspace{-0.1cm}
 {f_d}(x) =\frac{{d{F_d}(x)}}{{dx}} =  \frac{{3{x^2}}}{{{D^3} - {r^3}}},r \le x \le D.
\end{equation}

We adopt the 3-D channel model proposed in \cite{Hourani2014} which is more practical than the widely used free space channel model. This model captures the fact that the line-of-sight (LoS) probability increases with elevation angle. In particular, the LoS probability is given by
\vspace{-0.2cm}
\begin{equation}\label{dwerji}
\vspace{-0.2cm}
  {P_{{\rm{LoS}}}} = \frac{1}{{1 + a\exp \left( { - b\left( {\theta  - a} \right)} \right)}},
\end{equation}
where $a$ and $b$ are positive constants that depend  on the environment and the values are given in \cite{Hourani2014}. For a given location of UAV, the mean path loss is given by \cite{Hourani2014}:
\vspace{-0.1cm}
\begin{equation}\label{xswdff}
\vspace{-0.1cm}
  L(\theta ,d) = \frac{A}{{1 + a\exp \left( { - b\left( {\theta  - a} \right)} \right)}} + 20{\log _{10}}\left( {{d}} \right) + C,
\end{equation}
where $A$ and $C$ are constants given by $A=\eta _{{\rm{LoS}}}-\eta _{{\rm{NLoS}}}$ and $C=20{\log _{10}}\left( {\frac{{4\pi {f_c}}}{c}} \right) + {\eta _{{\rm{NLoS}}}}$, respectively. The $\eta _{{\rm{LoS}}}$ and $\eta _{{\rm{NLoS}}}$ (in dB) are respectively the losses
corresponding to the LoS and non-LoS links. In general, $\eta _{{\rm{NLoS}}}$ is much larger than $\eta _{{\rm{NLoS}}}$ due to the severe path loss of NLoS. $f_c$ is the carrier frequency (Hz), and $c$ is the speed of light (m/s).

Assume that the transmission power from GCS to UAV is fixed as $P$ and the noise power at UAV is denoted as ${\sigma ^2}$, the signal-to-noise ratio (SNR) at the UAV is given by
\vspace{-0.1cm}
\begin{equation}\label{jjgjngkk}
  \gamma (\theta,d)=\tilde C{d^{ - 2}}{e^{  \frac{{\tilde A}}{{1 + a\exp \left( { - b(\theta  - a)} \right)}}}}
\end{equation}
where  $\tilde A = -A\frac{{\ln 10}}{{10}}>0$ and $\tilde C = \frac{P}{{{\sigma ^2}}}{10^{ - \frac{C}{{10}}}}$.

For the control information delivery from the GCS to the UAV, the channel blocklength is small. According to \cite{Polyanskiy2010IT}, for a simple point-to-point system with finite blocklength $M$, to guarantee the minimum decoding error probability of $\varepsilon$, the achievable data rate $R$ (bits per channel use) can be approximately as
\vspace{-0.2cm}
\begin{equation}\label{weafrf}
\vspace{-0.15cm}
R(\gamma (\theta,d) )\! \!=\! {{{\log }_2}\!\left( {1 \!+\! \gamma (\theta,d) } \right) \!-\!\frac{1}{\text{ln} 2} \!\sqrt {\frac{V(\gamma (\theta,d))}{M}} {Q^{ - 1}}\!\!\left( \varepsilon  \right)},
\end{equation}
where  ${Q^{ - 1}}\left(\cdot  \right)$ is the inverse function of $Q\left( x \right) = \frac{1}{{\sqrt {2\pi } }}\int_x^\infty  {{e^{ - \frac{{{t^2}}}{2}}}} dt$, and $V(\gamma)$ is the channel dispersion that is a function of SNR $\gamma$ and is given by $V(\gamma)=1-(1+\gamma)^{-2}$ \cite{Polyanskiy2010IT}.  The second term in (\ref{weafrf}) can be regarded as the penalty on the data rate due to the short blocklength.

{In the following, we aim to derive the exact average achievable data rate (AADR) with fixed decoding error probability  $\varepsilon$ when the UAV is uniformly distributed in the space shown in Fig.~\ref{systemodel}, which is $\bar{R } = \mathbb{E}{\{R(\gamma (\theta,d))\}}$. If the transmission needs to be finished within $T_{\rm{max}}$ seconds and the system bandwidth is denoted as $B$, the total number of channel uses is given by $M=BT_{\rm{max}}$. Then the average number of bits that can be transmitted is calculated as $L=BT_{\rm{max}} \bar{R }$, which is equal to the packet size. Therefore, the AADR performance is vital, which can provide engineering insights for the design of the packet size.}

\vspace{-0.2cm}
\section{Average Achievable Data Rate}
{The average achievable data rate (AADR) is written as
\vspace{-0.1cm}
\begin{equation}\label{dewfr}
  \bar{R } = \int_r^D {\int_{\theta_{\rm{min}}}^{90} {R(\gamma (x,y)){f_{d,\theta }}(x,y)dy} } dx,
\end{equation}
where ${{f_{d,\theta }}(x,y)}$ is the joint PDF of $d$ and $\theta$. Since the UAV is randomly deployed in the restricted space in Fig.~\ref{systemodel}, the PDF of $\theta$ is given by ${f_\theta }(y) = {1 \mathord{\left/
 {\vphantom {1 {\left( {90 - {\theta _{{\rm{min}}}}} \right)}}} \right.
 \kern-\nulldelimiterspace} {\left( {90 - {\theta _{{\rm{min}}}}} \right)}}$. In addition, since $d$ and $\theta$ are independent, the joint PDF of $d$ and $\theta$ are given by ${f_{d, \theta }}(x,y) = {f_d}(x){f_\theta }(y)$, where ${f_d}(x)$ is given in (\ref{jijioijojio}). In the following, we first use the GCQ method to accurately approximate (\ref{dewfr}). Then, we derive its tight lower bound in closed form.}
\vspace{-0.4cm}
\subsection{{Approximate Expression}}

{One can see that (\ref{dewfr}) can be rewritten as
\vspace{-0.2cm}
\begin{equation}\label{der}
\vspace{-0.2cm}
\bar R = \frac{1}{{90 - {\theta _{{\rm{min}}}}}}\frac{3}{{{D^3} - {r^3}}}\int_r^D {\underbrace {\int_{{\theta _{{\rm{min}}}}}^{90} {{x^2}R(\gamma (y,x))dy} }_{{I_1}(x)}dx}.
\end{equation}
It is very difficult to find the closed-form expression of $I_1(x)$. We adopt the GCQ \cite{Abramowitz} to approximate it.  Define ${q_1}(x,y) = {x^2}R(\gamma (y,x))$,  we have
\vspace{-0.1cm}
\begin{eqnarray}
\!\!\!\! \!\! {I_1}(x)\!\!\!\!\! &=&\!\!\!\!\!\! \int_{{\theta _{{\rm{min}}}}}^{90}\!\! {{q_1}} (x,y)dy \\
   \!\!\!\! &\approx& \!\!\!\!\!  \frac{{90 - {\theta _{{\rm{min}}}}}}{2}\!\sum\limits_{i = 1}^{{N_1}}  {{l_i}}\! \cdot\!{q_1}\!\!\left(\!  {\frac{{90\! -\! {\theta _{{\rm{min}}}}}}{2}{t_i} \!+\! \frac{{90\! + \!{\theta _{{\rm{min}}}}}}{2},x} \!\!\right)\label{huhgtrt}
\end{eqnarray}
where $t_i$ is the $i$-th zero of Legendre polynomials, $N_1$ is the number of terms, $l_i$ is the Gaussian weight given by Table (25.4) of \cite{Abramowitz}.}

{Then, by substituting (\ref{huhgtrt}) into (\ref{der}),  one has
\vspace{-0.1cm}
\begin{equation}\label{eer}
\vspace{-0.1cm}
\bar{R } =\frac{3}{{2}}\frac{{{1}}}{{{D^3} - {r^3}}}   I_2
\end{equation}
where $I_2$ is given by
\vspace{-0.1cm}
\begin{equation}\label{kokok}
\vspace{-0.1cm}
  {I_2} = \int_r^D {\underbrace {\sum\limits_{i = 1}^{{N_1}} {{l_i}} {q_1}\left( {\frac{{90 - {\theta _{{\rm{min}}}}}}{2}{t_i} + \frac{{90 + {\theta _{{\rm{min}}}}}}{2},x} \right)}_{{q_2}(x)}} dx
\end{equation}
\vspace{-0.1cm}
By using the similar method, one can approximate ${I_2}$ as
\vspace{-0.1cm}
\begin{equation}\label{wee7}
I_2\approx \frac{D-r}{2}\sum_{i=1}^{N_2}k_i \cdot q_2\left(\frac{D-r}{2} g_i +\frac{D+r}{2}\right),
\end{equation}
where $g_i$, $N_2$ and $k_i$ are similarly defined as in (\ref{huhgtrt}).}

{Finally, by substituting (\ref{wee7}) into (\ref{eer}), we have
\begin{equation}\label{wee8}
\begin{aligned}
\bar{R }\approx \frac{3}{{4}}\frac{{{D-r}}}{{{D^3} - {r^3}}}\sum_{i=1}^{N_2}k_i \cdot q_2\left(\frac{D-r}{2} g_i +\frac{D+r}{2}\right).
\end{aligned}
\end{equation}}

\vspace{-0.7cm}
\subsection{{Lower Bound of AADR}}

{In the following, we aim to derive the lower bound (LB) of the AADR in closed form. To this end, we first introduce the following function
\vspace{-0.2cm}
\begin{equation}\label{jhjttypij}
\vspace{-0.2cm}
f(x) = \ln \left( {1 + \frac{1}{x}} \right) - q\sqrt {\frac{{2x + 1}}{{{{(x + 1)}^2}}}}, x>0
\end{equation}
where $q$ is a positive constant. Then, we have the following lemma:}

{\emph{\textbf{Lemma 1}}: $f(x)$ is always non-negative when $0<x\le g^{-1}(q)$, where $g^{-1}(q)$ is the inverse function of $g(x)$ which is defined in (\ref{dwffrf}).}

{\emph{\textbf{Proof}}: \upshape  Please refer to Appendix \ref{proofoflemma1}. \hfill\rule{2.7mm}{2.7mm}}

{In the following lemma, we show more properties of $f(x)$.}

{\emph{\textbf{Lemma 2}}: $f(x)$  is a decreasing and convex function when $0<x\le g^{-1}(q)$.}

{\emph{\textbf{Proof}}: \upshape  Please refer to Appendix \ref{proofoflemma2}.  \hfill\rule{2.7mm}{2.7mm}}

{Based on Lemma 2, we start to derive the LB of the AADR. In particular, $R(\gamma)$ \footnote{{For notation simplicity, $\gamma$ means $\gamma(\theta,d)$ in the following derivations.}} can be rewritten as
\vspace{-0.15cm}
\begin{equation}\label{dwsfrf}
\vspace{-0.2cm}
  R(\gamma ) = \frac{1}{{\ln 2}}f\left( {\frac{1}{\gamma }} \right)
\end{equation}
where $f(\cdot)$ is defined in (\ref{jhjttypij}) with ${q} = \frac{{{Q^{ - 1}}({\varepsilon })}}{{\sqrt M }}.$ Hence, if $\gamma  \ge {1 \mathord{\left/
 {\vphantom {1 {{g^{ - 1}}(q)}}} \right.
 \kern-\nulldelimiterspace} {{g^{ - 1}}(q)}}$, $f(\cdot)$ is a convex function. In our considered system, the minimum $\gamma$ is achieved when $d=D$ and $\theta  = 0$ based on (\ref{jjgjngkk}), and the minimum value is ${\gamma _{\min }} = \tilde C{D^{ - 2}}{e^{\frac{{\tilde A}}{{1 + a\exp \left( {ab} \right)}}}}$. This imposes the constraint for the maximum transmission distance between the UAV and the GCS, which is given by
 \vspace{-0.1cm}
 \begin{equation}\label{xdferf}
 \vspace{-0.1cm}
   D \le \sqrt {\tilde C{e^{\frac{{\tilde A}}{{1 + a\exp \left( {ab} \right)}}}}{g^{ - 1}}(q)}  \buildrel \Delta \over = {D_{\max }}
 \end{equation}
By using the parameters in the simulation section, the maximum distance ${D_{\max }}$ can be up to 56.4 km and 56.8 km for Dense Urban and Suburban scenarios \cite{bor2016efficient}, respectively, which are far beyond the control distance for UAV. As a result, condition (\ref{xdferf}) always holds in practice.  Then, $f(\cdot)$ is a convex function when $r\le d\le D$ and $\theta_{\rm{min}}\le \theta\le 90$. Then, we have
\vspace{-0.1cm}
\begin{equation}\label{dwefrf}
\vspace{-0.1cm}
  \mathbb{E}\{ R(\gamma )\}  \!\!=\!\! \frac{1}{{\ln 2}}\mathbb{E}\left\{ {f\left( {\frac{1}{\gamma }} \right)} \right\} \!\!\ge \!\! \frac{1}{{\ln 2}}f\left( \mathbb{E}{\left( {\frac{1}{\gamma }} \right)} \right) \buildrel \Delta \!\over =\!\! {R_{{\rm{lb}}}}.
\end{equation}
To obtain ${R_{{\rm{lb}}}}$, we only need to derive $\mathbb{E}{\left( {\frac{1}{\gamma }} \right)} $, which is much more tractable than directly deriving the exact AADR.}

{The expression of $\mathbb{E}{\left( {\frac{1}{\gamma }} \right)} $ is given by
\vspace{-0.15cm}
\begin{equation}\label{dqwew}
\vspace{-0.15cm}
 \begin{aligned}
 & \mathbb{E}\left( {\frac{1}{\gamma }} \right) = \frac{{{{3 \tilde C}^{ - 1}}}}{{\left(90-\theta_{\rm{min}}\right)\left( {{D^3} - {r^3}} \right)}}\cdot \\&
 \qquad \qquad\underbrace {\int_r^D {{x^4}dx} }_U\underbrace {\int_{\theta_{\rm{min}}}^{90} {{e^{ - \frac{{\tilde A}}{{1 + a\exp \left( { - b\left( {y - a} \right)} \right)}}}}dy} }_V.
 \end{aligned}
\end{equation}
where $U$ can be readily calculated as $U = \frac{1}{5}\left( {{D^5} - {r^5}} \right)$.
By using variable substitutions, (3.351.5) and (3.352.5) of \cite{Gradshteyn}, as well as some simple manipulations,  one can have
\vspace{-0.1cm}
\begin{equation}\label{dw}
\vspace{-0.1cm}
V = \frac{{\tilde A}}{b}\left( {T({y_1}) - T({y_2})} \right)
\end{equation}
where ${y_1} = \frac{{1 + a{\rm{exp}}( - b({\theta _{\rm{min}}} - a))}}{{\tilde A}}$, ${y_2} = \frac{{1 + a{\rm{exp}}( - b(90 - a))}}{{\tilde A}}$,   $T(x) =\frac{e^{-\tilde A} \text{Ei}\left(\tilde A-\frac{1}{x}\right)}{\tilde A}-\frac{\text{Ei}\left(-\frac{1}{x}\right)}{\tilde A} $ and $\text{Ei}(z) =-\int_{-z}^{\infty } \frac{e^{-t}}{t} \, dt$ \cite{Gradshteyn}. Then, one can obtain the lower bound of $\bar{R}$ by inserting (\ref{dqwew}) into (\ref{dwefrf}).}

\vspace{-0.1cm}
\section{Simulation}
\vspace{-0.1cm}
In this section, simulation results are presented to verify the correctness of our derived results. The simulation parameters are set as: $r= 250 $ m, $D= 400$ m, $P=-20$ dB, $\sigma^2 = -173 $ dBm/Hz, $B=1$ MHz, $f_c=2.5$ GHz, and $c=3\cdot 10^{8}$ m/s. Two scenarios are considered: dense urban and suburban. The values of the corresponding parameters can be found in \cite{bor2016efficient}. The minimum elevation angle $\theta_{\rm{min}}$ is set to be 45 and 30 for dense urban and suburban, respectively. This is reasonable since the former scenario has higher density of obstacles than the latter one. Four curves are plotted: 1) The average Shannon Capacity (`Shannon'), which is equal to $\mathbb{E}{\{ {{\log }_2}\left( {1 + \gamma (\theta,d) } \right)\}}$. The value is obtained by averaging over $10^4$ random UAV locations; 2) The approximate value of AADR obtained by using the  GCQ method (`Chebyshev'), which is given in (\ref{wee8}); 3) The simulated AADR (`Simulation'), which is equal to $\bar{R } = \mathbb{E}{\{R(\gamma (\theta,d))\}}$. Its value is  obtained by averaging over $10^4$ random UAV locations; 4) The lower bound of AADR (`LB') given in (\ref{dwefrf}).

\begin{figure}
	\centering
	\includegraphics[width=0.43\textwidth]{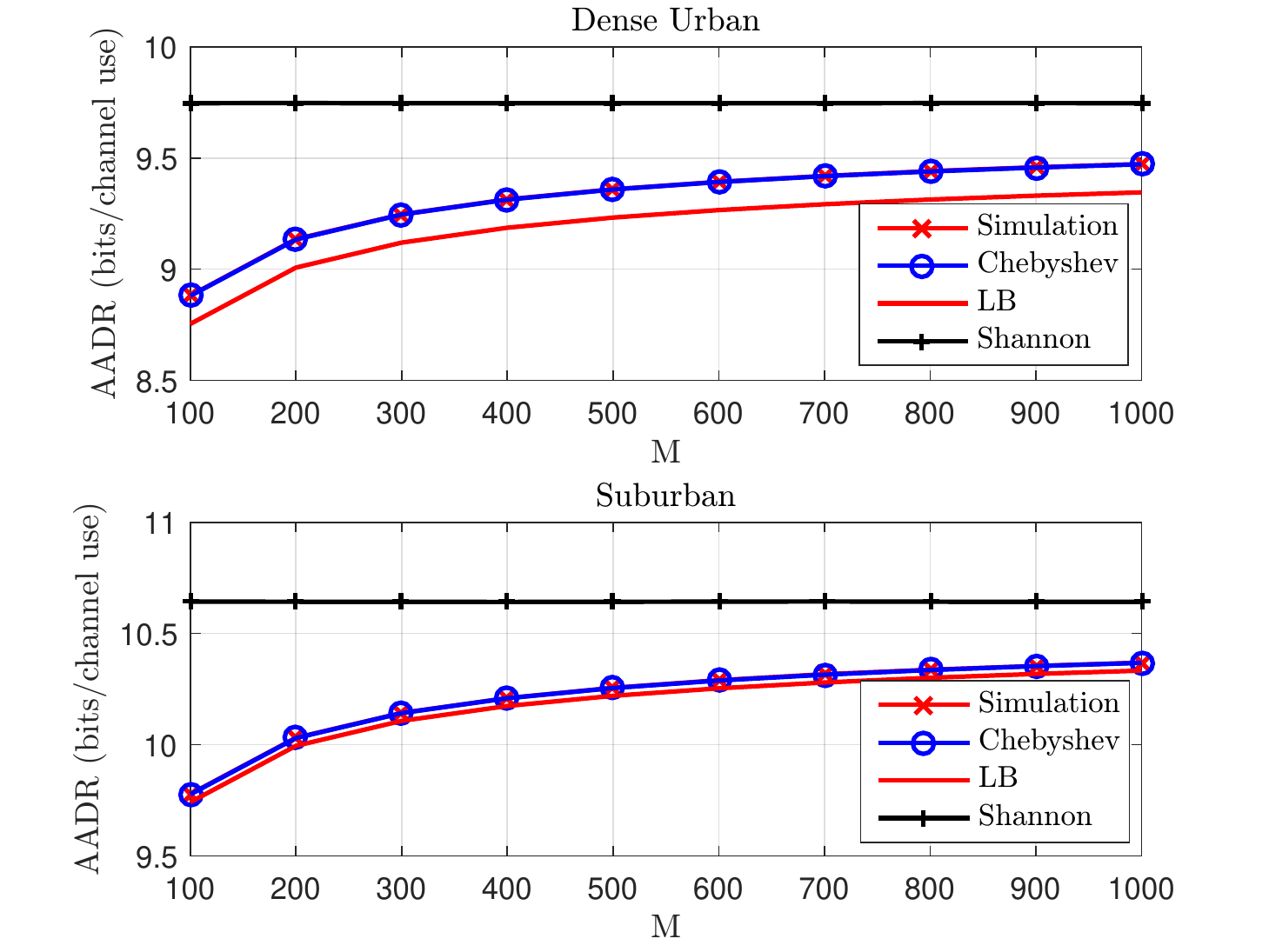}
    \vspace{-0.3cm}
	\caption{{AADR versus $M$ for two different scenarios.}}
	\vspace{-0.6cm}
	\label{fig1}
\end{figure}

In Fig.~\ref{fig1}, we first investigate the impact of channel blocklength $M$ on the AADR performance with the decoding error probability fixed at $\varepsilon=10^{-9}$. The channel blocklength $M$ ranges from 100 to 1000, and thus the corresponding latency ranges from 0.1 ms to 1 ms, which satisfies the low latency requirement.  From Fig.~\ref{fig1}, the AADR is observed to increase with  $M$ and approach the Shannon's capacity. The reason is that with the increase of $M$, the second term in (\ref{weafrf}) diminishes, which reduces to the Shannon's capacity formula. However, when $M$ is small, there is a big gap between the conventional Shannon's capacity and the AADR. This means that if directly adopting the Shannon's capacity as the performance measure, we will overestimate the system performance, which may cause unexpected incidents. Hence, we should adopt the short packet capacity to design the system.  In Fig.~\ref{fig1}, the results obtained by the GCQ method match very well with the simulation results. Furthermore, the gap between the `LB' and the `Simulation' is very small, especially in the case of the suburban scenario. These results verify the correctness of our derived results.

\begin{figure}
	\centering
	\includegraphics[width=0.43\textwidth]{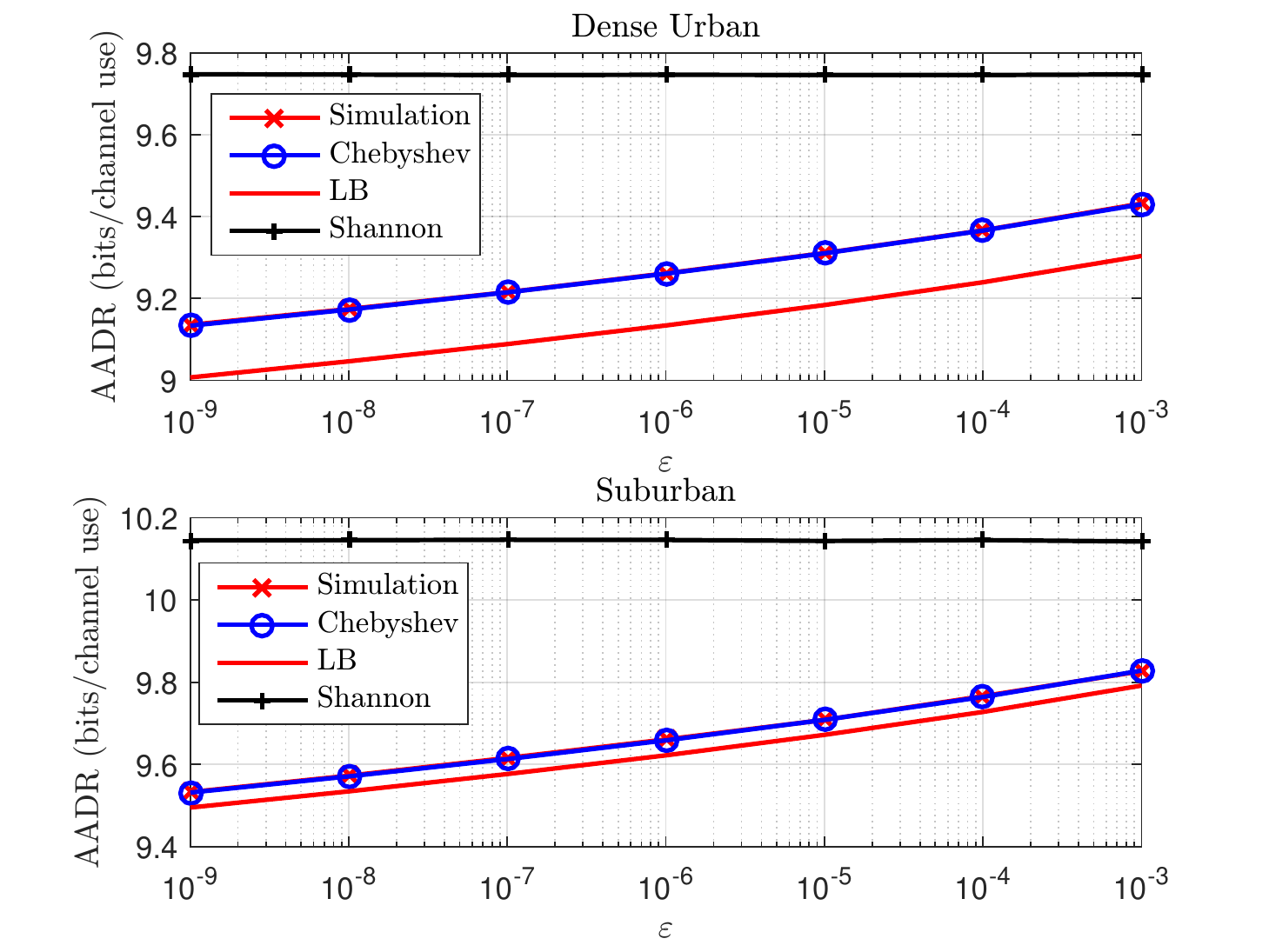}
    \vspace{-0.3cm}
	\caption{{AADR versus $\varepsilon$ for two different scenarios.}}
	\vspace{-0.6cm}
	\label{fig2}
\end{figure}

{In Fig.~\ref{fig2}, we investigate the impact of the decoding error probability $\varepsilon$ on the AADR performance where channel blocklength is set as $M=200$. It is seen from this figure that  the AADR is increasing with $\varepsilon$. This can be explained as follows: Function ${Q^{ - 1}}(x)$ is a monotonically decreasing function of $x$.  When the number of bits needed to transmit is large, the decoding error probability is high accordingly for both the same SNR and channel blocklength. There is big gap between the Shannon's capacity and the AADR, especially when $\varepsilon$ is very low. The numerical results obtained by the  GCQ method coincide with the simulated AADR, and the gap between LB and the simulated AADR is small especially when $\varepsilon$ is very low. This means that for highly reliable communications, the capacity under short channel blocklength should be adopted.
}

\vspace{-0.2cm}
\section{Conclusions}\label{conclu}

{In this letter, we have studied the AADR of a UAV communication system under short packet transmission using 3-D channel model. We have used the GCQ method to derive the approximate expression of the AADR. Then, a tight lower bound of AADR has been derived in closed form. These results are insightful for packet size design when guaranteeing a minimum decoding error probability target. Simulation results verify the correctness and the tightness of our derived results.}

\begin{appendices}
\vspace{-0.2cm}
\section{Proof of Lemma 1}\label{proofoflemma1}
{To guarantee that $f(x)$ is always non-negative, we must have
\begin{equation}\label{dwffrf}
q \le \frac{{(x + 1)\ln \left( {1 + \frac{1}{x}} \right)}}{{\sqrt {2x + 1} }} \buildrel \Delta \over = g(x).
\end{equation}
The first-order derivative of $g(x)$ with respect to $x$ is given by
\begin{equation}\label{dfeagtju}
g'(x) = \frac{{ - 2 - \frac{1}{x} + x\ln \left( {1 + \frac{1}{x}} \right)}}{{{{\left( {2x + 1} \right)}^{\frac{3}{2}}}}} \le \frac{{ - 1 - \frac{1}{x}}}{{{{\left( {2x + 1} \right)}^{\frac{3}{2}}}}} < 0
\end{equation}
where the second inequality follows by using $\ln \left( {1 + \frac{1}{x}} \right) < \frac{1}{x}$. Hence, $g(x)$ is a monotonically decreasing function of $x$. Then, the inequality in (\ref{dwffrf}) holds when $0<x\le g^{-1}(q)$, which completes the proof.}

\vspace{-0.2cm}
\section{Proof of Lemma 2}\label{proofoflemma2}
{The first-order derivative of $f(x)$ with respect to (w.r.t.) $x$ is given by
{\setlength\abovedisplayskip{3pt}
\setlength\belowdisplayskip{3pt}
\begin{equation}\label{joijijew}
f'(x) = \frac{1}{{1 + x}} - \frac{1}{x} - \frac{q}{{{{\left( {x + 1} \right)}^2}\sqrt {2x + 1} }} + \frac{{q\sqrt {2x + 1} }}{{{{\left( {x + 1} \right)}^2}}}.
\end{equation}}
\!\!To prove the monotonically decreasing property of $f(x)$, $f'(x)$ should be smaller than zero. Then, it should satisfy
{\setlength\abovedisplayskip{3pt}
\setlength\belowdisplayskip{3pt}
\begin{equation}\label{jorehjouth}
q < \frac{{(1 + x)\sqrt {2x + 1} }}{{2{x^2}}}\buildrel \Delta \over = {g_1}(x).
\end{equation}}
\!\!Next, we show that ${g_1}(x)>{g}(x)$ for any $x$:
\vspace{-0.2cm}
\begin{eqnarray}
\vspace{-0.2cm}
  {g_1}(x) &>& \frac{{(1 + x)\sqrt {2x + 1} \ln \left( {1 + \frac{1}{x}} \right)}}{{2x}} \label{jdhf}\\
    &>& \frac{{(1 + x)\sqrt {2x + 1} \ln \left( {1 + \frac{1}{x}} \right)}}{{2x + 1}} \\
    &=&  g(x)
\end{eqnarray}
where (\ref{jdhf}) follows by using the relation $\ln \left( {1 + \frac{1}{x}} \right) < \frac{1}{x}$. Since $0<x\le g^{-1}(q)$, we have $q \le g(x) < {g_1}(x)$. As a result, $f(x)$ is a monotonically decreasing function when $0<x\le g^{-1}(q)$.}
{The second-order derivative of   $f(x)$ w.r.t. $x$ is given by
{\setlength\abovedisplayskip{3pt}
\setlength\belowdisplayskip{3pt}\begin{equation}\label{pjpijpr}
f''(x) = \frac{{(x + 1){{(2x + 1)}^{\frac{5}{2}}} - 2q{x^2}\left( {3{x^2} - 1} \right)}}{{{x^2}{{(x + 1)}^3}{{(2x + 1)}^{\frac{3}{2}}}}}.
\end{equation}}
\!\!If $0 <  x\le {1 \mathord{\left/
 {\vphantom {1 {\sqrt 3 }}} \right.
 \kern-\nulldelimiterspace} {\sqrt 3 }}$, $f''(x)>0$ since $a$ is a positive value. Hence, $f(x)$ is a convex function for $0 <  x\le {1 \mathord{\left/
 {\vphantom {1 {\sqrt 3 }}} \right.
 \kern-\nulldelimiterspace} {\sqrt 3 }}$.  Next, we check the region $x > {1 \mathord{\left/
 {\vphantom {1 {\sqrt 3 }}} \right.
 \kern-\nulldelimiterspace} {\sqrt 3 }}$. In this case, to prove the convexity of $f(x)$, we need to show that $f''(x)>0$ when $x > {1 \mathord{\left/
 {\vphantom {1 {\sqrt 3 }}} \right.
 \kern-\nulldelimiterspace} {\sqrt 3 }}$, which yields
{\setlength\abovedisplayskip{3pt}
\setlength\belowdisplayskip{3pt}
\begin{equation}\label{itjhpihuorw}
q < \frac{{(x + 1){{(2x + 1)}^{\frac{5}{2}}}}}{{2{x^2}\left( {3{x^2} - 1} \right)}} \buildrel \Delta \over = {g_2}(x).
\end{equation}}
\!\!Next, we prove that ${g_2}(x)>g(x)$ as follows:
{\setlength\abovedisplayskip{3pt}
\setlength\belowdisplayskip{3pt}
\begin{equation}\label{jogsuhtruht}
\begin{array}{l}
{g_2}(x) = \frac{{x + 1}}{{x\sqrt {2x + 1} }}\frac{{{{(2x + 1)}^3}}}{{2x(3{x^2} - 1)}} > \\
\frac{{x + 1}}{{x\sqrt {2x + 1} }}\frac{{{{(2x + 1)}^3}}}{{6{x^3}}} > \frac{{x + 1}}{{x\sqrt {2x + 1} }} > g(x).
\end{array}
\end{equation}}
\!\!Hence, ${g_2}(x)>q$ and $f(x)$ is also a convex function over $x > {1 \mathord{\left/
 {\vphantom {1 {\sqrt 3 }}} \right.
 \kern-\nulldelimiterspace} {\sqrt 3 }}$.
}
{Based on the above two cases, function $f(x)$ is a convex function over the entire region when  $0<x\le g^{-1}(q)$.}

\end{appendices}

\
\

\vspace{-0.4cm}
\bibliographystyle{IEEEtran}
\bibliography{myre}


\end{document}